\documentclass[letterpaper, 10 pt, conference]{ieeeconf} 
\IEEEoverridecommandlockouts  
\usepackage[utf8]{inputenc}
\usepackage{amsmath,amssymb,euscript ,yfonts,psfrag,latexsym,dsfont,graphicx,bbm,color,amstext,wasysym,subfig,flushend,parskip}
\usepackage{times,mathptmx,epsfig,graphicx,lineno,color}

\definecolor{grey}{rgb}{0.6,0.6,0.6}
\definecolor{lightgray}{rgb}{0.97,.99,0.99}

\definecolor{llgrey}{rgb}{0.9,0.9,0.9}
\definecolor{lgrey}{rgb}{0.6,0.6,0.6}
\definecolor{lred}{rgb}{0.9,0.7,0.7}

\newcommand{\black}{\color{black}}
\definecolor{dblue}{rgb}{0,0,.5}

\usepackage{xcolor}
\modulolinenumbers[5]

\newtheorem{prop}{Proposition}

\newtheorem{remark}{Remark}

\newcommand{\trace}{{\rm trace\,}}
\newcommand{\cQ}{Q}
\newcommand{\cW}{W}

\def\spacingset#1{\def\baselinestretch{#1}\small\normalsize}
\spacingset{1}

\begin{document}

\title{Minimal entropy production in anisotropic temperature fields}

\author{Olga Movilla Miangolarra$^\star$, Amirhossein Taghvaei$^\dagger$, and Tryphon T. Georgiou$^\star$
\thanks{$^\star$Mechanical and Aerospace Engineering, University of California, Irvine, CA 92697, USA;
omovilla@uci.edu, tryphon@uci.edu}
\thanks{$^\dagger$Aeronautics and Astronautics Department, University of Washington, Seattle, Washington 98195, USA; amirtag@uw.edu}}

\maketitle

\begin{abstract}
Anisotropy of temperature fields, chemical potentials and ion concentration gradients provide the fuel that feeds dynamical processes that sustain life. Dynamical flows in respective environments incur losses manifested as entropy production. In this work we consider a rudimentary model of an overdamped stochastic thermodynamic system in an anisotropic temperature heat bath, and analyze the problem to minimize entropy production while driving the system between thermodynamic states in finite time.
It is noted that entropy production in a fully isotropic temperature field, can be expressed as the Wasserstein $W_2$ length of the path traversed by the thermodynamic state of the system. In the presence of an anisotropic temperature field, the mechanism of entropy production is substantially more complicated as, besides dissipation, it entails seepage of energy between the ambient heat sources by way of the system dynamics. We show that, in this case, the entropy production can be expressed as the solution of a suitably constrained and generalized Optimal Mass Transport (OMT) problem. In contrast to the situation in standard OMT, entropy production may not be identically zero, even when the thermodynamic state remains unchanged. Physically, this is due to the fact that maintaining a Non-Equilibrium Steady State (NESS), incurs an intrinsic entropic cost. As already noted, NESSs are the hallmark of life and living systems by necessity operate away from equilibrium. Thus our problem of minimizing entropy production appears of central importance in understanding biological processes, such as molecular motors and motor proteins, and on how such processes may have evolved to optimize for available usage of resources.
\end{abstract}


\section{Introduction}

Life on Earth is possible thanks to the temperature gradient between the hot Sun and the cold stary sky. This anisotropy in thermal excitation where photons are absorbed at around 6000 Kelvin and emitted back to the cosmos, twenty fold at a dramatically reduced temperature of about 300 Kelvin, provides the ``negative entropy'' that organisms and complex biochemical processes feed upon \cite{opatrny2017life}.
It is precisely the anisotropy in the thermal environment that sustains dynamical flows and non-equilibrium steady states that make life possible \cite{battle2016broken,gnesotto2018broken}.
The ``hard currency'' paid along the way, in accordance with the second law of thermodynamics, is the entropy of the universe that keeps increasing. 

In the present work we analyze the propensity of a thermodynamic system to increase entropy via a rudimentary model that captures the effects of an anisotropic temperature field. Our interest is in quantifying the minimal amount of entropy production during thermodynamic transitions in anisotropic temperature fields, and casting optimization of such as a suitably constrained stochastic control problem.

It is of essence to highlight that, in spite of the centrality of the notion entropy in thermodynamics since the foundational work of Carnot and Clasius \cite{carnot1824reflections}, it was not until recently, within the framework of stochastic thermodynamics, that the entropy and energy budget during finite-time thermodynamic transitions could be explicitly quantified. Indeed, it has been a recent discovery that for mesoscopic systems modeled via Langevin stochastic differential equations, dissipation can be expressed via the Wasserstein $W_2$ length that thermodynamic states traverse during transition \cite{aurell2011optimal,dechant2019thermodynamic,chen2019stochastic}.
These developments have opened up new directions of research pertaining to the design of optimal control protocols \cite{fu2021maximal,abiuso2022W2Carnotthermodynamics}, quantifying speed limits and establishing uncertainty relations \cite{nakazato2021geometrical,van2022thermodynamic,ito2022geometric},
and developing geometric formalisms for the emerging theoretical framework
\cite{EnergyHarvestingAnisotropic2021,Dechant2021geometric,miangolarra2022geometry}. Within this evolving landscape the study of entropy production in anisotropic temperature fields remains largely unexplored.


The structure of the present paper is as follows. Section \ref{sec:stochasticthermo} summarizes the basic framework that allows quantitative description of finite-time transitions. Section \ref{sec:control} discusses challenges in the presence of an anisotropic temperature field, followed by analysis for the case where the control authority includes non-conservative forces \ref{sec:nonconservative}, to be contrasted in Section \ref{sec:constrained} with constrained stochastic control problems that arise when the control authority is restricted to a gradient of a controlling potential.

\section{Stochastic Thermodynamic Systems}\label{sec:stochasticthermo}

Stochastic Thermodynamics \cite{sekimoto2010stochastic,seifert2012stochastic,peliti2021stochastic} has been conceived to model thermodynamic processes that evolve both in discrete as well as in continuous state space, utilizing Master Equations or Langevin Stochastic Differential equations, respectively; herein we restrict our attention to the latter.

A thermodynamic system, at a mesoscopic scale, can be conceptualized as a collection of particles in contact with heat baths, modeled as sources of stochastic excitation, while driven under the influence of external forces. These forces can be conservative (gradients) or non-conservative and the dynamics may or may not include inertial effects. In the present work we focus on overdamped dynamics (that is, we do not consider inertial effects). Such models are typical when considering colloidal mesoscopic particle systems and models of biological processes. Most importantly, the systems we consider are in contact with multiple heat baths. Thus, our basic model is the Langevin system
 \begin{equation}\label{eq:Langevin}
dX_t=-\gamma^{-1}\nabla U(t,X_t) dt+ \gamma^{-1}f(t,X_t)dt+ \sqrt{2D}dB_t 
 \end{equation}
where $X_t\in \mathbb R^n$, $t\in\mathbb R$ representing time, $\nabla U(t,X_t) dt$ representing the conservative forces of the drift term (typically constituting our control), while $f$ represents the non-conservative ones, and the diffusion tensor $D$ abiding by the Einstein relation
\[
D = k_B \gamma^{-1} T,
\]
with $T$ a diagonal matrix with entries the value of temperature (in Kelvin) along the specified $n$ degrees of freedom, and $\gamma$ a scalar friction coefficient. While there is not difficulty in pursuing the analysis in this generality, the essence already comes through when we consider the case $n=2$.

The state of the thermodynamic system is represented by the probability density function $\rho(t,x)$ that satisfies the Fokker-Planck (FP) equation
\begin{equation}\label{eq:FP}
\partial_t \rho(t,x) +\nabla \cdot J(t,x) =0,
\end{equation}
with $x=(x_1,\ldots,x_n)^\prime\in\mathbb R^n$, and probability current
\begin{align}\label{eq:pcurrent}
J&=-\gamma^{-1}
\left(
\rho\nabla U - f + k_BT\nabla \rho
\right) = \rho v(t,x).
\end{align}
As is common, $\nabla := \left(\begin{matrix}\partial_1, &\ldots, & \partial_n\end{matrix}\right)^\prime$, with $\partial_i:=\frac{\partial}{\partial x_i}$, denotes the gradient and ``$\nabla\cdot\;\;$'' the divergence.
 For the most part (with the exception of Section \ref{sec:nonconservative}), we will assume the absence of non-conservative forces, i.e., that $f=0$.
Note that $v(t,X_t)$ represents a velocity field in Lagrangian coordinates. Another set of convenient notations are
\[
\beta:=\frac{1}{k_B\sqrt[n]{\det(T)}}, \mbox{ and }R:= -\beta^{-1}\log(\rho).
\]
In this way, $\rho=e^{-\beta R}$ while
\[v=-\gamma^{-1}(\nabla U +\mathbf T\nabla R),
\]
with $R$ a ``stochastic potential'' and $\mathbf T=T/\sqrt[n]{\det(T)}$ a normalized temperature tensor. Unless the temperature is isotropic, the velocity field $v$ fails to be the gradient of a potential, and thereby  generates circulating currents.

\subsection{The first law}
The rate of change of the internal energy 
\[
\mathcal E=\mathbb E\{U(t,X)\}=\int \rho(t,x) U(t,x)dx,
\]
namely,\footnote{Throughout, $dx$ is a short for the volume form $dx_1\dots dx_n$.}
\[
\frac{d\mathcal E}{dt} = \underbrace{\int \rho \dot Udx}_{\dot\cW}+\underbrace{\int \dot \rho Udx}_{\dot\cQ},
\]
splits into the part where the system exchanges work with the external potential $U$, and the part where energy is exchanged with the heat bath(s). Accordingly, $\dot\cW$ {\em represents the rate of work done to the system} by way of changing the potential $U$, while
\begin{align*}
\int \dot \rho Udx &= - \int U \nabla\cdot J dx= -\sum_{i=1}^n\int U \partial_i J_i dx,
\end{align*}
decomposes the heat that flows into the system into the contributions from the different sources. Specifically,
 \[
\dot \cQ= \sum_i \dot \cQ_i,
\]
with $\dot \cQ_i = -\int U \partial_i J_i dx$ the heat drawn from the $i$th reservoir.

\subsection{The second law}
The total entropy production includes two terms, entropy production within the system and entropy increase in the environment, namely,
\[
\dot S_{\rm tot} = \dot S_{\rm sys}+\dot S_{\rm env}
\]
where (using integration by parts)
\[
\dot S_{\rm sys}=-k_B\int \dot \rho\log\rho dx= -k_B\int \langle J, \nabla \log\rho\rangle dx
\]
and
\[
\dot S_{\rm env}=-\sum_i \frac{\dot \cQ_i}{T_i}=-\sum_i\frac{-\int U\partial_i J_idx}{T_i} =-\int \langle J,T^{-1}\nabla U\rangle dx.
\]
The minus sign is due to the convention that (positive) heat-rate $\dot\cQ_i$ {\em is taken out of the environment and into the system}.
Thus, together, 
\begin{align*}
\dot S_{\rm tot}&=-
\int \langle J, T^{-1} \left(\nabla U + k_BT\nabla \log\rho\right)\rangle dx\\
&=\gamma \int \frac{1}{\rho}\|J\|^{2}_{ T^{-1}}dx,
\end{align*}
where we have used Eq. \eqref{eq:pcurrent}.
Therefore, any non-zero probability current $J$ that effects a thermodynamic transition or helps maintain a non-equilibrium steady state, irreversibly increases the total entropy, as per the second law of thermodynamics.
Alternatively, one can express the entropy production rate in terms of $v$ as 
\begin{align}
\dot S_{\rm tot}=\int \rho\|v\|^{2}_{\gamma T^{-1}}dx.\label{eq:entropyproduction}
\end{align}

\section{Control authority and dissipation cost} \label{sec:control}

Thermodynamic states are assumed to have finite variance. Their respective space\footnote{Herein, probability distributions are assumed absolutely continuous
with respect to the Lebesgue measure and thus represented by density functions; see \cite[Sec.\ 8]{ambrosio2005gradient} for the general case.} is denoted by $\mathcal P_2(\mathbb R^n)$ (or, $\mathcal P_2$ for simplicity). Interestingly, this space admits a very rich structure that renders it almost a Riemannian manifold \cite{ambrosio2005gradient}. Much of what follows to a large degree can be traced to this.

At any given $\rho\in \mathcal P_2$, the tangent space of $\mathcal P_2$ can be thought of as the space of admissible perturbations $\rho\to \rho+\delta$ for suitable $\delta(x)$'s that integrate to zero.  It can be shown that these ``tangent directions'' $\delta$ can be placed in bijective correspondence with gradient vector fields $v=\nabla \phi$ satisfying the Poisson equation $\delta=-\nabla\cdot(\rho\nabla \phi)$.
Then, the inner product $\int \rho \langle \nabla \phi_1,\nabla \phi_2\rangle dx$ allows computing length of paths between densities -- the smallest distance (geodesic) between any given $\rho_0$ and $\rho_f$ is known as the Wasserstein metric $W_2(\rho_0,\rho_f)$; see \cite[Remark 8.4]{villani2003topics} and \cite[Section 8.4]{ambrosio2005gradient} for a detailed exposition. 

\subsection{Control authority}
Thus, returning to the FP equation 
\eqref{eq:FP},
$\partial_t \rho +\nabla \cdot \rho v =0$, 
the available control authority by manipulating the potential $U(t,x)$ in
\begin{align}\label{eq:vform}
v&=-(\gamma^{-1}\nabla U+D\nabla \log \rho)
\end{align}
is enough to specify the gradient part in $v$, and thereby, any tangent direction $\delta$. 
Specifically, in order to move in a given tangent direction $\delta$, the control $U$
 has to satisfy the following Poisson equation
\[
\nabla\cdot(\rho \nabla U)
=\gamma\delta- \nabla\cdot (k_B T\nabla \rho).
\]
It follows that
$
U=U_\rho +\delta_U
$
where $U_\rho$ and $\delta_U$ solve 
\[
\nabla\cdot(\rho \nabla U_\rho)
=- \nabla\cdot k_B T\nabla \rho,\mbox{ and }
\nabla\cdot(\rho \nabla \delta_U)
=\gamma\delta,
\]
respectively, for the specified $\rho,\gamma,T$ and $\delta$.
Note that both equations have a unique solution.

Thus, controllability of \eqref{eq:FP} when prescribing the controlling potential $U$ is precisely the same as that when $v$ is completely unconstrained. Equation \eqref{eq:vform} simply reveals that a portion of $v$ is annihilated by the divergence operator, while it still contributes to the entropy production. Indeed, by invoking the
Helmholtz' decomposition of vector fields, we highlight below the significance of the constituent terms in $v$.

\subsection{Helmholtz' decomposition of vector fields}

We endow vector fields on $\mathcal X=\mathbb R^n$ with inner product  
\[
\langle v_1,v_2\rangle_{\rho,M}= \int \rho(x) \langle v_1(x),v_2(x)\rangle_{M} dx,
\]
for any $\rho\in\mathcal P_2$ and underlying Euclidean inner-product $\langle v_1(x),v_2(x)\rangle_{M}=v_1(x)^\prime Mv_2(x)$, for a symmetric positive-definite matrix $M$. Then,
\begin{equation}\label{eq:orthogonal}
v=M^{-1}\nabla \phi + \frac{\chi}{\rho} ,
\end{equation}
with $\chi$ a solenoidal (i.e., divergence-free $\nabla\cdot \chi=0$) vector field,
is an orthogonal decomposition.
To see this note that $\langle M^{-1}\nabla \phi,\frac{\chi}{\rho}\rangle_{\rho,M}=0$ for all $\phi$, via integration by parts assuming differentiability and suitably fast decay at infinity.

A solenoidal vector field $\chi$ in $\mathbb R^3$ can be expressed via the curl of a vector potential $A$ as $\chi=\nabla\times A$. In $\mathbb R^2$ this reduces to
$
\chi=\Omega\nabla \psi
$
where, $\Omega$ is a skew-symmetric matrix (i.e., so that $\Omega+\Omega^\prime=0$).
Thus, e.g., specializing to 
$\mathbb R^2$, for the above orthogonal decomposition \eqref{eq:orthogonal},
\begin{equation}\label{eq:orthogonal7}
\|v\|^2_{\rho,M}= \int \rho \|\nabla\phi\|^2_{M^{-1}} +\int \frac{\|\nabla \psi\|^2_{\Omega^\prime M \Omega }}{\rho}.
\end{equation}



\section{Non-conservative actuation}\label{sec:nonconservative}

It is rather instructive to consider entropy production under the full authority of non-conservative actuation. That is, with control actuation that entails both a gradient $\nabla U$ of a potential as well as a non-zero term $f$ in \eqref{eq:Langevin} contributing with a solenoidal component. In this case, the minimal dissipation turns out to coincide with a suitably weighted Wasserstein length traversed by the thermodynamic state. This fact represents a geometric characterization of entropy production and highlights a link between thermodynamics of overdamped dynamics and optimal mass transport \cite{aurell2011optimal}, albeit the thermal anisotropy in our case necessitates a suitable re-weighing of the optimal transport cost as explained below.

\subsection{Dissipation as a weighted Wasserstein length}
{The least entropy production over paths $\rho(t,\cdot)$ between end-point states,
using \eqref{eq:entropyproduction}, 
is precisely the least weighted transport cost
\begin{align}\label{eq:dynamicW2}
    W_{2,M}^2(\rho_0,\rho_{f}) &:=\min_{\rho,v} \ \int_0^{1}\int \rho \|v\|^2_M dxdt\\
    &= t_f \min_{\rho,v}\int^{t_f}_0\dot S_{\rm tot}\, dt \nonumber,
\end{align}
where $M=\gamma T^{-1}$ and the optimization is subject to  $\partial_t\rho+\nabla\cdot (\rho v)=0$ together with the end-point conditions $\rho(0)=\rho_0$ and $\rho(t_f)=\rho_f$.} 
We relate this {\em weighted Wasserstein distance} $W_{2,M}$ to an un-weighted (corresponding to the identity matrix as weight) Wasserstein distance as follows.

First we invoke the fact that an optimal transportation plan requires constancy of the velocity along paths (in Lagrangian view point); this follows by the Cauchy-Schwartz
inequality. Thus, for a mass element (particle) that starts at location $x$ and terminates at $y$ over the time interval $[0,t_f]$, the velocity remains constant and equal to \begin{align}\label{eq:v}
    v(X(x,t),t)=(y-x)/t_f
\end{align}
with path traversed $X(x,t)=x+tv$ for $t\in[0,t_f]$.
This standard observation turns the dynamic optimal transport \eqref{eq:dynamicW2} into a static (Kantorovich-type) problem, so as to be subsequently cast as an unweighted transport problem via a change of variables, as follows. Specifically, let
$\pi$ be a distribution on the product space $(x,y)\in\mathbb R^n\times \mathbb R^n$ that represents the law of pairing origin $x$ to destination $y$, under a transport policy. Thus $\pi$ is a coupling of random variables $X(x,0)$ and $X(y,t_f)$, with probability density functions $\rho_0(x)$ and $\rho_{t_f}(y)$, respectively; these are marginal distributions of $\pi$ and this is the only condition for $\pi$ to be a ``coupling.'' Then,
\begin{align*}
    W_{2,M}^2 &= \min_{\pi} \int \|x-y\|_M^2 d \pi 
\\&= 
\min_{\pi} \int \|M^{\frac{1}{2}}x-M^{\frac{1}{2}}y\|^2 d \pi 
\\&=W_2^2( M^{\frac{1}{2}}\# \rho_0, M^{\frac{1}{2}}\# \rho_{t_f})
\end{align*}
where, with a slight abuse of notation, $M^{\frac{1}{2}}\# \rho_0$ denotes the push-forward with the map $x\mapsto M^{\frac{1}{2}}x$.
Using standard theory \cite{villani2003topics}, the optimal transport map for unweighted transport is given by the gradient of a convex function $\varphi$, and hence we now have $x\mapsto y =M^{-\frac{1}{2}} \nabla \varphi(M^{\frac{1}{2}} x)$; here, $\nabla \varphi$ is the optimal transport map between $M^{\frac{1}{2}}\# \rho_0$ and $M^{\frac{1}{2}}\# \rho_f$ for unweighted cost. 

One can also express the weighted metric as
\begin{align*}
    W_{2,M}^2(\rho_0,\rho_f) = \frac{\gamma}{\sqrt[n]{\det(T)}}W_2^2( \mathbf{T}^{-\frac{1}{2}}\# \rho_0, \mathbf{T}^{-\frac{1}{2}}\# \rho_{t_f}),
\end{align*}
where $\mathbf{T}=\frac{T}{\sqrt[n]{\det(T)}}$ is the normalized temperature tensor given earlier as a volume preserving transformation. A geometrical procedure to find the optimal transportation is to "warp" the space according to $\mathbf{T}$, identify the optimal transport in the usual way, and then "warp"  back.

\subsection{Dissipation for Gaussian thermodynamic states.}

In general, for standard optimal mass transport problems, explicit solutions are hard to come by and need to be computed numerically. The same of course applies to the case of weighted transport.
One exception is when transport traces paths on the submanifold of Gaussian distributions. In such cases the Wasserstein-2 distance can be written explicitly.

For completeness, we provide here the weighted Wasserstein-2 distance between two normal distributions:
\begin{align*}W_{2,M}(\rho_0,\rho_{t_f})&=\Big[\|\mu_x-\mu_y\|^2_M+ \trace \big\{\Sigma_x M+\Sigma_yM\\&-2\Sigma_xM\Sigma_y^{1/2}(\Sigma_y^{1/2}M\Sigma_x M\Sigma_y^{1/2})^{-1/2}
\Sigma_y^{1/2}M\big\}\Big]^{1/2},
\end{align*}
where $\rho_0=\mathcal N(\mu_x,\Sigma_x)$ and $\rho_{t_f}=\mathcal N(\mu_y,\Sigma_y)$.
The derivation can be carried out as in the unweighted case \cite{bhatia2019bures}.

\section{Dissipation cost under conservative actuation} \label{sec:constrained}

We now consider the case where our control is limited to conservative forcing, i.e., $f=0$. We first solve the local problem of steering a state $\rho$ in the direction of minimal entropy production, and then consider optimal transitioning between endpoint distributions keeping the entropy production minimal.

\subsection{Direction minimizing entropy production}
We are interested in identifying a potential $U$ that minimizes the rate of entropy production, locally. That is, we want to characterize a potential that steers the thermodynamic state in a direction where the rate of entropy production is the smallest possible.

\begin{prop} For any given $\rho\in\mathcal P_2$, a necessary condition for a potential $U$ to minimize the local entropy production rate $\dot S_{\rm tot}=\int \|v\|^2_M \rho dx$, where $v=-\gamma^{-1}(\nabla U+k_BT\nabla \log \rho)$ and $M=\gamma T^{-1}$, is
\begin{equation}\label{eq:tomatch1}
    \nabla\cdot ( \rho T^{-1}\nabla U)= - k_B\Delta \rho.
\end{equation}
\end{prop}

\begin{proof}
The first variation of $\dot S_{\rm tot}$ with respect to the controlling potential $U$ is
\begin{align*}
 & 2\int \langle \nabla \delta_U,\gamma^{-1}T^{-1}\rho(\nabla U+k_BT\nabla\log\rho )\rangle dx =\\&  
 -2\int   \delta_U \nabla\cdot\big(\gamma^{-1}T^{-1}\rho(\nabla U+k_BT\nabla\log\rho )\big)dx ,
\end{align*}
using integration by parts.
Setting the variation to zero for all perturbations $\delta_U$, we readily derive \eqref{eq:tomatch1} as first-order optimality condition.
\end{proof}
 Note that $\nabla U$ can be interpreted as the projection of $-k_BT \nabla \log(\rho)$ into the space of gradient vector-fields with respect to $\langle{\cdot,\cdot}\rangle_{\rho,T^{-1}}$, i.e.
\begin{equation*}
   U= {\rm arg}\min_U \| \nabla U + k_BT\nabla \log(\rho) \|^2_{\rho,T^{-1}}.
\end{equation*}

\begin{remark}
{
To ascertain existence of a solution to \eqref{eq:tomatch1}, we may express this as the Poisson equation 
\[
\mathcal L_{\rho}  U = h,
\] 
where $\mathcal L_{\rho}  U := -\frac{1}{\rho} \nabla \cdot(\rho T^{-1}\nabla  U)$ is the (weighted) Laplacian operator and $h:=\frac{1}{\rho}k_B \Delta p$. Let $L^2_\rho$ denote the space of square-integrable functions with respect to $\rho$, equipped with the inner-product $\langle f,g\rangle_\rho:= \int fg\rho dx$, and $H^1_\rho$ denote the Sobolev space of functions whose first derivatives (in the weak-sense) are in $L^2_\rho$. Note that $\mathcal L_{\rho}$ is a positive symmetric operator with respect to the $L^2_\rho$ inner-product, and that it has a trivial eigenvalue at $0$, with constants as eigenfunctions. Existence and uniqueness of a (weak) solution is guaranteed provided that  $\|h\|_\rho^2$ is finite and $\mathcal L_\rho$ satisfies a spectral gap condition, i.e.,    
\begin{equation*}
   \langle f,\mathcal L_\rho f\rangle_\rho \geq \lambda \|f\|_\rho^2,
 \end{equation*}
 for some $\lambda>0$ and all functions $f\in H^1_\rho$ orthogonal to constants (i.e. $\int f \rho dx = 0$)~\cite{laugesen2015poisson}. The spectral gap condition is equivalent to the Poincar\'e inequality for $\rho$, which holds under mild assumptions for $\rho$, e.g. a Gaussian tail~\cite{bakry2014analysis}.

}
\end{remark}

\subsection{Minimal entropy production between end-points}

We now consider conditions for minimizing entropy production along a path between two endpoint distributions for velocity fields constrained as explained in Section \ref{sec:control}.

We first rewrite the rate of entropy production as
\begin{align*}
   \dot S_{\rm tot} =&\gamma^{-1} \int \|\nabla U+k_BT\nabla \log \rho\|^2_{ T^{-1}}\rho dx\\=& \gamma^{-1} \Big[\int \|\nabla U\|^2_{T^{-1}} \rho dx-k_B^2 \int \|\nabla \log \rho\|^2_T\rho dx\Big]+2\dot S_{\rm sys},
\end{align*}
using that $\dot S_{\rm sys}=-k_B\int \partial_t\rho\log\rho dx=-k_B\int (\nabla\log\rho)'v\rho dx$ via integration by parts. Since $\int_0^{t_f}\dot S_{\rm sys}dt=S_{\rm sys}(\rho(t_f))-S_{\rm sys}(\rho(0))$ only depends on the endpoint distributions, minimizing entropy production over the transition amounts to solving
\begin{equation}\label{eq:equiv-cost}
    \min_{U,\rho} \gamma^{-1} \int_0^{t_f} \int\Big[ \|\nabla U\|^2_{ T^{-1}} -k_B^2\|\nabla \log \rho\|^2_T\Big]\rho dxdt,
\end{equation}
subject to the continuity equation \eqref{eq:FP} and the endpoint conditions. Necessary conditions for optimality are stated below.

\begin{prop} A path $\rho(t,\cdot)$  between specified terminal states, along with the corresponding control protocol $U(t,\cdot)$ that solve
\begin{align*}
\min_{U,\rho} \Big\{  \int_{0}^{t_f}\dot S_{\rm tot} dt
    \mid \eqref{eq:FP} \mbox{ and } \rho(0)=&\rho_0,\ \rho(t_f)=\rho_f\Big\},
\end{align*}
satisfy
\begin{subequations}\label{eq:FONC}
\begin{align}\label{eq:FONCa}
\gamma \partial_t\rho  =&\nabla\cdot(\rho(\nabla U+k_BT\nabla\log\rho))\\ \nonumber
  \gamma\partial_t\lambda=&\|\nabla U\|^2_{T^{-1}} 
  +\tfrac{2k_B^2}{\rho}\nabla\cdot(T \nabla \rho)
 -k_B^2\|\nabla \log \rho\|^2_T
   \\\label{eq:FONCb}
   &+\langle\nabla\lambda,\nabla U+ k_BT\nabla \log \rho\rangle
  -\tfrac{1}{\rho}\nabla\cdot(\rho k_BT\nabla\lambda)\\
  0=& \nabla\cdot(2T^{-1} \rho\nabla U
  +\rho\nabla\lambda)\label{eq:FONCc}.
\end{align}
\end{subequations}
\end{prop}

\begin{proof}
We use the expression in \eqref{eq:equiv-cost} to write the following augmented Lagrangian:
\begin{align*}
    J=&\gamma^{-1} \int_0^{t_f} \int\Big[ (\nabla U)^\prime T^{-1} \nabla U-k_B^2(\nabla \log \rho)^\prime T \nabla \log \rho\Big]\rho dx dt\\&+\int_0^{t_f} \int\lambda\Big[\partial_t\rho-\gamma^{-1}\nabla\cdot((\nabla U+k_BT\nabla\log\rho)\rho)\Big]dxdt,
\end{align*}
with $\lambda$ a Lagrange multiplier. The first variation is
\begin{align*}
    \delta J=&\gamma^{-1}\int_0^{t_f} \int\Big[ 2(\nabla \delta_U)^\prime T^{-1} \nabla U\rho+(\nabla U)^\prime T^{-1} \nabla U\delta_\rho\\
    &-2k_B^2\big(\nabla \tfrac{\delta_\rho}{\rho}\big)^\prime T \nabla \log \rho\rho-k_B^2(\nabla \log \rho)^\prime T \nabla \log \rho\delta_\rho\\
    &+\lambda\big\{\gamma\partial_t\delta_\rho-\nabla\cdot((\nabla U+k_BT\nabla\log\rho)\delta_\rho)\\
    &-\nabla\cdot((\nabla \delta_U+k_BT\nabla\tfrac{\delta_\rho}{\rho})\rho)\big\}\\
    &+\delta_\lambda\big(\gamma\partial_t\rho-\nabla\cdot((\nabla U+k_BT\nabla\log\rho)\rho)\big)\Big] dxdt.
\end{align*}
Integrating by parts and setting this to zero for all perturbations $\delta_U,\delta_\rho,\delta_\lambda$ we obtain \eqref{eq:FONC}.
\end{proof}

\begin{remark}
In contrast to the minimal entropic cost of Section \ref{sec:nonconservative}, the entropy production here, where the control is restricted to being a gradient of a potential, is no longer a distance between end-point states. This is evident since  maintaining a stationary state may require non vanishing entropy production.
\end{remark}

\begin{remark}
The equations \eqref{eq:FONC} need, in general, to be solved numerically, e.g., by iterating while solving (\ref{eq:FONCa}-\ref{eq:FONCb}) forward in time with $U$ computed via \eqref{eq:FONCc}.
\end{remark}

\black

\subsection{Geometric decomposition of entropy production}\label{sec:geometric}

Consider a given trajectory $\rho(t,\cdot)\in \mathcal P_2$ that connects end-point states $\rho_0,\rho_f$, and let $\delta:=\partial_t\rho$. The gradient part $v_{\rm grad}:= M^{-1}\nabla \phi$ of any velocity field $v$ (as in \eqref{eq:orthogonal})
that realizes this trajectory is fixed -- the divergence-free part impacts entropy production but not the evolution of the state.
Thus, the entropy production $\int_0^{t_f}\int \|v\|_{\gamma T^{-1}}^2\rho dxdt$ can be decomposed as
\begin{equation}\label{eq:decomp}
    \int_0^{t_f}\int \|v_{\rm grad}\|_{\gamma T^{-1}}^2\rho dxdt +
\int_0^{t_f}\int \|\frac{\chi}{\rho}\|_{\gamma T^{-1}}^2\rho dxdt,
\end{equation}
analogously to \eqref{eq:orthogonal7}.

The first of these two contributions represents the minimal entropy production that is attainable when we allow non-conservative actuation (cf.\ Section \ref{sec:nonconservative}) -- it is precisely the Wasserstein action integral for the space equipped with the $\langle \cdot,\cdot \rangle_{\gamma T^{-1}}$ Riemannian metric.  Thus, it constitutes a lower bound to the total entropy production \eqref{eq:decomp}. It can be thought of the entropic cost related to the steering of the thermodynamic state.

The second term in \eqref{eq:decomp}
represents a contribution to the entropy production that is due to circulation in the velocity field. Such circulation is needed to sustain a non-equilibrium steady state (NESS). For this reason, this contribution to entropy production has been referred to as ``housekeeping entropy production''.
The  decomposition of entropy production in \eqref{eq:decomp} can be seen as a generalization to anisotropic temperature fields of analogous decompositions presented in \cite{nakazato2021geometrical,Dechant2021geometric,ito2022geometric}.

\begin{remark}
In the above, $\phi$ is obtained by solving
\begin{align*}
\nabla\cdot(\rho\gamma^{-1}T\nabla \phi)=-\delta.
\end{align*}
The controlling potential $U$ is obtained so that $\nabla U$ is the gradient part of $-T\nabla(k_B\log \rho+\phi)$. Finally, the divergence-free component $\chi$ is obtained from
\begin{align*}
\frac{\chi}{\rho}=-\gamma^{-1}(\nabla U+T\nabla(k_B\log \rho+\phi)).
\end{align*}
\end{remark}

\black
\section{Control via a quadratic potential}
We now specialize to the case where the controlling potential is quadratic, namely,
$
U(t,x)= x^\prime  K(t) x/2
$
with $x\in \mathbb R^n$ and $K(t)=K(t)^\prime >0$. The thermodynamic state traces a path on the submanifold of Gaussian distributions
\[
\rho(t,x)=
\frac{1}{(2\pi)^{n/2}\det(\Sigma(t))^{1/2}} e^{-\frac12 \|x\|^2_{\Sigma(t)^{-1}}},
\]
where the covariance $\Sigma$ satisfies the differential Lyapunov equation (corresponding to \eqref{eq:FP})
\begin{equation}\label{eq:Lyapunov} \tag{2$^\prime$}
    \gamma \dot \Sigma = -K\Sigma -\Sigma K +2k_BT.
\end{equation}
Since $\nabla \log\rho= -\Sigma^{-1}x$, from \eqref{eq:vform},
$
v(t,x)=-\gamma^{-1} Kx +D\Sigma^{-1}x.
$

We use these expressions to find the optimal control, and therefore the optimal $K(t)$, that steers the system into the {\em direction of minimal entropy production}. Specifically, from  \eqref{eq:tomatch1} we obtain the following necessary condition for optimality:
\begin{equation}\label{eq:tomatch2} \tag{10$^\prime$}
    T^{-1} K \Sigma+  \Sigma K T^{-1} =2k_B I,
\end{equation}
equivalently, 
\[
\tilde K\tilde \Sigma +\tilde \Sigma\tilde K=2k_BT,
\]
where $\tilde K=T^{-1/2}KT^{-1/2}$ and $\tilde \Sigma=T^{1/2}\Sigma T^{1/2}$. This is an algebraic Lyapunov equation. Its solution can be written in closed form as
\[
K=2k_B\int_0^\infty T^{1/2} e^{-T^{1/2}\Sigma T^{1/2}\tau}Te^{-T^{1/2}\Sigma T^{1/2}\tau}T^{1/2}d\tau,
\]
so that the corresponding choice of $U$ minimizes entropy production locally in time.

Next we specialize the first-order optimality conditions \eqref{eq:FONC} for {\em transition between end-point states} to Gaussian states and transition path. We adopt the ansatz that the Lagrange multiplier is of the form
$$
\lambda(t,x)=\frac12 x'\Lambda(t)x+c(t).
$$
The optimal $K,\Sigma$ and $\Lambda$ satisfy
\begin{subequations}
\begin{align}\tag{12${\rm a}^\prime$}
\gamma\dot\Sigma=&-K\Sigma-\Sigma K+2k_BT,\\\tag{12${\rm b}_1^\prime$}
\gamma\dot\Lambda=&\Lambda K+K\Lambda+2KT^{-1}K+2k_B^2\Sigma^{-1}T\Sigma^{-1},\\
\tag{12${\rm b}_2^\prime$}
\gamma \dot c=&-2k_B^2\,\trace(T\Sigma^{-1})-k_B\,\trace[T\Lambda],\\
   0=& 2T^{-1}K\Sigma+2\Sigma K T^{-1}+\Lambda\Sigma+\Sigma\Lambda,\tag{12${\rm c}^\prime$}
\end{align}
\end{subequations}
translating \eqref{eq:FONC} to the quadratic actuation case.
This is a set of coupled algebraic-differential equations with two-point boundary conditions, that can be solved numerically in a way analogous to the general case (via a shooting method).

Lastly, we specialize the {\em geometric decomposition} of entropy production of Section \ref{sec:geometric} to the case of a Gaussian path of distributions. Assuming zero-mean, the path amounts to a curve of covariance matrices $\{\Sigma(t):t\in[0,t_f]\}$. The entropy production is now given by
\begin{align*}
\int_0^{t_f}\dot S_{\rm tot}dt=\int_0^{t_f}\trace[X' \gamma T^{-1}X]dt, 
\end{align*}
where $X=-\gamma^{-1} K\Sigma^{1/2} +D\Sigma^{-1/2}$ can be decomposed as 
$$
X=X_{\rm s}+X_{\rm a},
$$
with $X_{\rm s}= \gamma^{-1} TA$ and $A$ the symmetric matrix that renders $X-X_{\rm s}$ anti-symmetric.
Then,
$A$ solves the Lyapunov equation
$$
AT^{-1}+T^{-1}A=
\Big[-\Sigma^{1/2}K-K\Sigma^{1/2}+k_B\Sigma^{-1/2}T+k_BT\Sigma^{-1/2}\Big].
$$
Accordignly, thanks to the orthogonality condition $\trace[X_{\rm s}{\gamma T^{-1}}X_{\rm a}]=0$, the entropy production decomposes into two parts
$$
\int_0^{t_f}\trace[X_{\rm s}{\gamma T^{-1}}X_{\rm s}]dt+\int_0^{t_f}\trace[X_{\rm a}{\gamma T^{-1}}X_{\rm a}]dt,
$$
in agreement with \eqref{eq:decomp}.

\section{Conlcusions}

In these pages we have attempted to grasp intrinsic characteristics of entropy production in an anisotropic (temperature) setting, especially its idiosyncratic unavoidability. We have highlighted the necessity of non-conservative forcing when attempting to stall the entropy production associated to anisotropy. We have characterized the directions in which entropy production increases the least, and provided conditions that optimal paths must satisfy to minimize entropy production for transitioning between two endpoint distributions. Yet, a lot remains to be understood, such as conditions pertaining to the existence of optimal controls, robustness to uncertainty in the constituents and environment, and so on. We believe that this endeavour, ultimately, is of prime importance for understanding of the interplay between entropy production and mechanisms in biology that might be essential for life.

\bibliography{mybibfile}

\begin{thebibliography}{10}
\providecommand{\url}[1]{#1}
\csname url@samestyle\endcsname
\providecommand{\newblock}{\relax}
\providecommand{\bibinfo}[2]{#2}
\providecommand{\BIBentrySTDinterwordspacing}{\spaceskip=0pt\relax}
\providecommand{\BIBentryALTinterwordstretchfactor}{4}
\providecommand{\BIBentryALTinterwordspacing}{\spaceskip=\fontdimen2\font plus
\BIBentryALTinterwordstretchfactor\fontdimen3\font minus
  \fontdimen4\font\relax}
\providecommand{\BIBforeignlanguage}[2]{{%
\expandafter\ifx\csname l@#1\endcsname\relax
\typeout{** WARNING: IEEEtran.bst: No hyphenation pattern has been}%
\typeout{** loaded for the language `#1'. Using the pattern for}%
\typeout{** the default language instead.}%
\else
\language=\csname l@#1\endcsname
\fi
#2}}
\providecommand{\BIBdecl}{\relax}
\BIBdecl

\bibitem{opatrny2017life}
T.~Opatrn{\`y}, L.~Richterek, and P.~Bakala, ``Life under a black sun,''
  \emph{American Journal of Physics}, 2017.

\bibitem{battle2016broken}
C.~Battle, C.~P. Broedersz, N.~Fakhri, V.~F. Geyer, J.~Howard, C.~F. Schmidt,
  and F.~C. MacKintosh, ``Broken detailed balance at mesoscopic scales in
  active biological systems,'' \emph{Science}, 2016.

\bibitem{gnesotto2018broken}
F.~Gnesotto, F.~Mura, J.~Gladrow, and C.~P. Broedersz, ``Broken detailed
  balance and non-equilibrium dynamics in living systems: a review,''
  \emph{Reports on Progress in Physics}, 2018.

\bibitem{carnot1824reflections}
S.~Carnot, ``Reflections on the motive power of fire, and on machines fitted to
  develop that power,'' \emph{Paris: Bachelier}, 1824.

\bibitem{aurell2011optimal}
E.~Aurell, C.~Mej{\'\i}a-Monasterio, and P.~Muratore-Ginanneschi, ``Optimal
  protocols and optimal transport in stochastic thermodynamics,''
  \emph{Physical review letters}, vol. 106, 2011.

\bibitem{dechant2019thermodynamic}
A.~Dechant and Y.~Sakurai, ``Thermodynamic interpretation of {W}asserstein
  distance,'' \emph{arXiv preprint arXiv:1912.08405}, 2019.

\bibitem{chen2019stochastic}
Y.~Chen, T.~Georgiou, and A.~Tannenbaum, ``Stochastic control and
  non-equilibrium thermodynamics: fundamental limits,'' \emph{IEEE Transactions
  on Automatic Control}, vol.~65, no.~1, pp. 252--262, 2020.

\bibitem{fu2021maximal}
R.~Fu, A.~Taghvaei, Y.~Chen, and T.~T. Georgiou, ``Maximal power output of a
  stochastic thermodynamic engine,'' \emph{Automatica}, vol. 123, p. 109366,
  2021.

\bibitem{abiuso2022W2Carnotthermodynamics}
P.~Abiuso, V.~Holubec, J.~Anders, Z.~Ye, F.~Cerisola, and M.~Perarnau-Llobet,
  ``Thermodynamics and optimal protocols of multidimensional quadratic brownian
  systems,'' \emph{arXiv preprint arXiv:2203.00764}, 2022.

\bibitem{nakazato2021geometrical}
M.~Nakazato and S.~Ito, ``Geometrical aspects of entropy production in
  stochastic thermodynamics based on wasserstein distance,'' \emph{Physical
  Review Research}, 2021.

\bibitem{van2022thermodynamic}
T.~Van~Vu and K.~Saito, ``Thermodynamic unification of optimal transport:
  Thermodynamic uncertainty relation, minimum dissipation, and thermodynamic
  speed limits,'' \emph{arXiv preprint arXiv:2206.02684}, 2022.

\bibitem{ito2022geometric}
S.~Ito, ``Geometric thermodynamics for the fokker-planck equation: Stochastic
  thermodynamic links between information geometry and optimal transport,''
  \emph{arXiv preprint arXiv:2209.00527}, 2022.

\bibitem{EnergyHarvestingAnisotropic2021}
O.~Movilla~Miangolarra, A.~Taghvaei, R.~Fu, Y.~Chen, and T.~T. Georgiou,
  ``Energy harvesting from anisotropic fluctuations,'' \emph{Phys. Rev. E}, Oct
  2021.

\bibitem{Dechant2021geometric}
A.~Dechant, S.-i. Sasa, and S.~Ito, ``Geometric decomposition of entropy
  production in out-of-equilibrium systems,'' \emph{Phys. Rev. Res.}, Mar 2022.

\bibitem{miangolarra2022geometry}
O.~Movilla~Miangolarra, A.~Taghvaei, Y.~Chen, and T.~T. Georgiou, ``Geometry of
  finite-time thermodynamic cycles with anisotropic thermal fluctuations,''
  \emph{arXiv preprint arXiv:2203.12483}, 2022.

\bibitem{sekimoto2010stochastic}
K.~Sekimoto, \emph{Stochastic energetics}.\hskip 1em plus 0.5em minus
  0.4em\relax Springer, 2010, vol. 799.

\bibitem{seifert2012stochastic}
U.~Seifert, ``Stochastic thermodynamics, fluctuation theorems and molecular
  machines,'' \emph{Reports on progress in physics}, 2012.

\bibitem{peliti2021stochastic}
L.~Peliti and S.~Pigolotti, \emph{Stochastic Thermodynamics: An
  Introduction}.\hskip 1em plus 0.5em minus 0.4em\relax Princeton University
  Press, 2021.

\bibitem{ambrosio2005gradient}
L.~Ambrosio, N.~Gigli, and G.~Savar{\'e}, \emph{Gradient flows: in metric
  spaces and in the space of probability measures}.\hskip 1em plus 0.5em minus
  0.4em\relax Springer Science \& Business Media, 2005.

\bibitem{villani2003topics}
C.~Villani, \emph{Topics in optimal transportation}.\hskip 1em plus 0.5em minus
  0.4em\relax American Mathematical Soc., 2003, no.~58.

\bibitem{bhatia2019bures}
R.~Bhatia, T.~Jain, and Y.~Lim, ``On the bures--wasserstein distance between
  positive definite matrices,'' \emph{Expositiones Mathematicae}, 2019.

\bibitem{laugesen2015poisson}
R.~S. Laugesen, P.~G. Mehta, S.~P. Meyn, and M.~Raginsky, ``Poisson's equation
  in nonlinear filtering,'' \emph{SIAM Journal on Control and Optimization},
  2015.

\bibitem{bakry2014analysis}
D.~Bakry, I.~Gentil, M.~Ledoux \emph{et~al.}, \emph{Analysis and geometry of
  Markov diffusion operators}.\hskip 1em plus 0.5em minus 0.4em\relax Springer,
  2014, vol. 103.

\end{thebibliography}
\bibliographystyle{IEEEtran}

\end{document}